# Decoherence-free subspaces in supersymmetric oscillator networks

Jeffrey Satinover[1]

Department of Physics, Yale University, New Haven, Connecticut

ABSTRACT: If protected against decoherence superpositions can be used for quantum information processing. A two-particle four-state system may have two-dimensional subspaces that are partially decoherence-free (e.g.., a symmetric triplet state), or completely so (e.g., an anti-symmetric singlet state.) A multi-particle system that in the laboratory basis is plagued by decoherence may in some other basis exhibit the symmetries that yield such decoherence-free subspaces (DFS's). Fully-interacting many-fermion spin-½ networks may be mathematically transformed to a more tractable many-to-one (or –to-some) variant. This paper applies such a transformation to a hypothetical network of boson-like operators and then argues that a fully-interacting particle number-preserving network of bosons plus fermions with supersymmetric degrees of freedom may be similarly be exploited so as to contain DFS's.

Decoherence is the major obstacle to the implementation of practical quantum information processing. Decoherence arises when the relative phases between component kets in an ensemble of single-particle superpositions become uncorrelated. However, the relative phase between the component kets of the maximally-entangled, anti-symmetric two-particle, two-state singlet state vector

$$|\psi\rangle = \frac{|01\rangle - |10\rangle}{\sqrt{2}} \qquad (1)$$

is stationary under the influence of interactions either with an external environment of similar two-state particles—no matter how many—or between the two particles themselves, regardless of the basis of action or the basis of representation [1, 2].

Within the four-dimensional Hilbert space spanned by the basis $\{|00\rangle, |01\rangle, |10\rangle, |11\rangle\}$, the two-dimensional subspace spanned by $\{e^{i\phi t}(+\frac{1}{\sqrt{2}}|01\rangle, -\frac{1}{\sqrt{2}}|10\rangle)\}$ is a decoherence-free subspace (DFS). The resistance of the subspace to decoherence (variation in the relative phase of its two component kets) arises from its internal anti-symmetry: external interactions can induce no phase difference, only a common overall, hence ignorable phase. $|\psi\rangle$ as expressed in Eq. (1) is effectively a two-particle eigenstate of the interaction Hamiltonian for each of two particles and an ensemble of many other particles, e.g.:

$$\frac{\hbar}{2}\sum_{i=2}^{N} \omega_{0i} \sigma_0 \sigma_i^+ + \frac{\hbar}{2}\sum_{i=2}^{N} \omega_{1i} \sigma_1 \sigma_i^+ \qquad (2)$$

Other states are partially resistant to decoherence. The symmetric triplet state:

$$|\psi\rangle = \frac{|01\rangle + |10\rangle}{\sqrt{2}} \qquad (3)$$

is fully decoherence-free in a $\sigma_z$-acting environment, but fully decohered by a $\sigma_x$-acting environment.

A physical instance of the above is a two-spin-½-particle system affected by an environment of any number of other spin-½ particles acting in any number of different bases, but without intra-system or intra-environment interaction.

---

[1] jeffrey.satinover@yale.edu    http://pantheon.yale.edu/~satinovr

However, anything short of full interaction among all particles is technically difficult to implement. For the singlet state of Eq. (1), this will have no importance. But for the triplet state of Eq. (3), an effectively randomized basis of interaction will ensure that the system state will become thoroughly decohered.

In a previous paper [1], a simplified Bogoliubov transform was applied to an ensemble of fully-interacting particles so as to represent it in two different diagonalized bases, one for the environment and one for the system. The resulting Hamiltonian is precisely of the form of Eq. (2). One value of such a transform is in greatly simplifying the complexity of numerical simulations of decoherence, from exponential in the number of spins to polynomial. If all the coupling constants are known (for example, when all elements of the environment act upon every element of the system in the same fashion), this transformation may be additionally useful in allowing us to extract an otherwise hidden symmetry from a highly asymmetrical-appearing ensemble. If the coupling constants are arbitrary however, it then turns out that they may be ignored only when in the original basis the target system is in the singlet state of Eq. (1). Such a state is not only inherently decoherence-free, it is also invariant under a change of basis. Hence we know how to express it in any basis whatever, and knowledge of the coupling constants that determine the change-of-basis transformation is superfluous.

That the anti-symmetry of the state in Eq. (1) renders it both decoherence-free and invariant are two ways of saying the same thing, the first expressed as an action upon the state, the second as a change of the basis of representation.

There is yet a third helpful way of understanding why anti-symmetrical states may resist decoherence. A purely dephasing environment of many spins (i.e., each environmental spin acts exclusively in the *z*-basis upon a *z* system spin, $\sigma_z$) may be thought of as imparting to the system spin a sequence of "phase-kicks" [2], particle by particle. If the system spin is in the state $|0\rangle$, then the net kick at time $\tau$ may be expressed as:

$$e^{i\varphi(\tau)} = \prod_{j=1}^{N-1} e^{i\varphi_j(\tau)} \tag{4}$$

But if the system spin is in the state $|1\rangle$, then the net kick at time $\tau$ is:

$$e^{-i\varphi(\tau)} \tag{5}$$

For a system spin in either of the two eigenstates $|0\rangle$ or $|1\rangle$, the factors Eq. 0.5 or Eq. 0.5 are irrelevant. But if the system spin is in some superposed state $\alpha|0\rangle + \beta|1\rangle$, then the phase kicks for each of the two component kets are opposite, not common, and the result is a change in phase between the components apart from any common phase factor:

$$\alpha|0\rangle + \beta|1\rangle \xrightarrow{\tau} \alpha e^{i\varphi(\tau)}|0\rangle + \beta e^{-i\varphi(\tau)}|1\rangle = e^{-i\varphi(\tau)}\left(\alpha e^{2i\varphi(\tau)}|0\rangle + \beta|1\rangle\right) \tag{6}$$

However, if the system is the two particle singlet state of Eq. (0.1), the environment acts identically on the two (entangled) two component kets $|01\rangle$ and $|10\rangle$, even though they are in some sense "opposites." The imparted phase kick is common to both, hence ignorable, as in the case of single-particle eigenstates. Note that, for a network of spin-½ particles, the fact that the energy eigenvalues are equal and opposite is not what produces a DFS; Rather, it is that we have arranged to superpose (more precisely, "entangle") two distinguishable states in such a way that they are at once "symmetrical" (in having the same energy), yet "opposite" (in phase) in a way that is not attainable using any single-particle superposition.

The importance of this anti-symmetry leads naturally to an investigation of *supersymmetrical* structures as perhaps also affording protection against decoherence. The simplest supersymmetrical multiparticle system is composed of single two-state fermion and a single multi-state boson. We first explore some specifics of a purely bosonic network. We may then examine a supersymmetric network.

A network of bosonic oscillators, any two of its (exclusively positively eigenvalued) energy eigenstates may be similarly entangled so as to form a protected DFS. Of course, since the bosonic oscillator has not only two but an infinite set of discrete energy levels, the possible structures of DFS's, even within a mere two-particle Hilbert space, can be significantly more complex, and there are infinitely more such structures.

For simplicity, (but anticipating what will happen naturally anyway in a full-fledged supersymmetric Hamiltonian), we temporarily replace the full bosonic Hamiltonian:

$$\frac{\hbar\omega}{2}(b^+b+\mathbf{1}) \to \frac{\hbar\omega}{2}(b^+b) \tag{7}$$

(as though the ground state energy was exactly zero; or as though we were describing an exchange interaction only). The Hamiltonian for a fully-interacting network of N such boson-like oscillators may be written:

$$H_b = \frac{\hbar}{2}\sum_{i=0}^{N}\sum_{j=0}^{N}\omega_{ij}\left(b_i^+ \otimes b_j \prod \otimes \mathbf{1}_k\right) = \frac{\hbar}{2}\sum_{i=0}^{N}\sum_{j=0}^{N}\omega_{ij}b_i^+b_j = \frac{\hbar}{2}\sum_{i=0}^{N}\sum_{j>i}^{N}\left(\omega_{ij}b_i^+b_j + \omega_{ij}^*b_ib_j^+\right) \tag{8}$$

where $b_i^+$ and $b_i$ respectively represent the creation and annihilation operators for the $i^{th}$ boson site in the network. There are no restrictions on the interaction strength $\omega_{ij}$; $\mathbf{1}$ is the identity matrix for the $k^{th}$ Hilbert space.

For a fermionic network, a term in the Hamiltonian such as $\sigma_i\sigma_j$ (= $\sigma_i^+\sigma_j$) enters the propagator in the exponential $e^{im_im_jt}$ as the product of the two respective energy eigenvalues $m_i = \pm 1$ and $m_j = \pm 1$, depending on the state vectors of the $i^{th}$ and $j^{th}$ spins, i.e.,

$$(+1_i)(+1_j) = +1 \qquad\qquad (+1_i)(-1_j) = -1$$
$$(-1_i)(+1_j) = -1 \qquad\qquad (-1_i)(-1_j) = +1$$

We may therefore express the propagator directly from the Hamiltonian.

But for bosons, $n = 0, +1, +2, \ldots$ are not the eigenvalues of the creation/annihilation operators; they are eigenvalues of the number operator $b_i^+b_i$.

This crucial asymmetry of bosons vis-à-vis fermions can be highlighted in an especially helpful way by comparing the form of the bosonic network Hamiltonian of Eq. (8) with the fermionic equivalent:

$$H_f = \frac{\hbar}{2}\sum_{i=0}^{N}\sum_{j=0}^{N}\omega_{ij}\left(\sigma_i^+ \otimes \sigma_j \prod \otimes \mathbf{1}_k\right) \tag{9}$$

In the fermionic Hamiltonian, the spin-operator product element $\sigma_i^+\sigma_j$ plays the same role as the bosonic number-operator product element $b_i^+b_i$. But $\sigma$ and $\sigma^+$ are not the fermionic equivalents of $b$ and $b^+$, rather $f$ and $f^+$ are, with

$$\sigma = f^+ f - \frac{\mathbf{1}}{2} \qquad (10)$$

where **1** is the identity operator and for a spin-½ system we may explicitly represent $f$, $f^+$ and $\sigma$ as:

$$f = \begin{pmatrix} 0 & 0 \\ 1 & 0 \end{pmatrix}, f^+ = \begin{pmatrix} 0 & 1 \\ 0 & 0 \end{pmatrix}, f^+ f = \begin{pmatrix} 1 & 0 \\ 0 & 0 \end{pmatrix}, f^+ f - \frac{\mathbf{1}}{2} = \begin{pmatrix} \tfrac{1}{2} & 0 \\ 0 & -\tfrac{1}{2} \end{pmatrix} = \frac{\sigma_z}{2} \qquad (11)$$

We have therefore:

$$\sigma_i^+ \sigma_j = \left(f_i^+ f_i - \tfrac{1}{2}\right) \otimes \left(f_j^+ f_j - \tfrac{1}{2}\right) = f_i^+ f_i f_j^+ f_j - \tfrac{1}{2}\left(f_i^+ f_i + f_j^+ f_j\right) + \tfrac{1}{4} \qquad (12)$$

Thus, we cannot write the fermionic Hamiltonian in the ladder-operator formalism in the algebraic form as its bosonic counterpart, viz.:

$$H_f \neq \frac{\hbar}{2} \sum_{i=0}^{N} \sum_{j=0}^{N} \omega_{ij} f_i^+ f_j = \frac{\hbar}{2} \sum_{i=0}^{N} \sum_{j>i}^{N} \left(\omega_{ij} f_i^+ f_j + \omega_{ij}^* f_i f_j^+\right) \qquad (13)$$

Nonetheless, we can apply a simplified Bogoliubov transformation independently to the two different representations of the Hamiltonian: to the spin-operator product representation for fermions and to the ladder-operator product representation for bosons, yielding a transformed basis of non-interacting fermions and non-interacting bosons, respectively.

Recapitulating the transformation in [1] but this time using bosonic creation and annihilation operators in place of spin operators and setting $\hbar = 2$, we may rewrite the scalar $H_b$ in Eq. (0.9) in terms of an N × N Hermitian coefficient matrix **H** and the N-dimensional vectors of annihilation operators and of creation operators:

$$H_b = \mathbf{b} \mathbf{H} \mathbf{b}^+ \qquad (14)$$

All (or any subspace) of **H** may be diagonalized:

$$H_b = \mathbf{b}^+ \mathbf{H} \mathbf{b} = \mathbf{b}^+ \left(\mathbf{U} \mathbf{H_D} \mathbf{U}^+\right) \mathbf{b} = \left(\mathbf{b}^+ \mathbf{U}\right) \mathbf{H_D} \left(\mathbf{U}^+ \mathbf{b}\right) = \mathbf{b}'^+ \mathbf{H_D} \mathbf{b}' \qquad (15)$$

Each original particle thus contributes to a new basis of quasi-particles represented by a set of quasi-ladder-operators $b_j'$. The Hamiltonian is that for N non-interacting (quasi-)bosons. To simplify the modeling of decoherence induced in a "system" by an "environment" (as in [1]), one performs this transformation independently on all of the "system" particles and all of the "environmental" ones: The interaction between system and environment remains unchanged, but the internal interactions within each will have vanished.

The particular two-particle anti-symmetric superposition of both possible spin-½ kets expressed in Eq. (1) is a coherent "eigenstate." Similarly, there exist multi-particle superpositions of all possible boson kets that are like "eigenstates" of the bosonic creation and annihilation operators—"coherent superpositions" as they are appropriately termed. So, to obtain (by definition) a DFS for a bosonic network, one might simply generate a coherent state. But this requires superposing in just the right proportion not merely two kets, as with spins, but an infinity of energy kets. While such states clearly exist in nature—indeed they are a consequence of decoherence itself [3]—they present severe obstacles to any finite-state approach to information processing.

On the other hand, if we restrict our computational basis to any finite number of bosonic states (two being the simplest), we will have difficulty both in deriving the propagator in the algebraic (ladder) representation because the kets are not eigenkets of the creation (or annihilation) operators, and in creating good approximations to a DFS. Only when the number of particles is very large may we consider *n* a good approximation to the correct eigenvalue; and only when the number of states in a superposition is large may we approximate a coherent state.

The reduction of a fully-interacting Hamiltonian to the non-interacting form is therefore of great value—at least for obtaining exact solutions. (We take up shortly its value with respect to DFS's.) First, each of the component state vectors for all the individual bosons constituting system and environment can be re-expressed in terms of operators acting on the vacuum state. Thus for a bosonic network of N+1 particle-positions:

$$|\Psi\rangle = |n_0 n_1 n_2 \ldots n_N\rangle = \left( \frac{(b_0^+)^{n_0}}{\sqrt{n_0!}} \frac{(b_1^+)^{n_1}}{\sqrt{n_1!}} \frac{(b_2^+)^{n_2}}{\sqrt{n_2!}} \cdots \frac{(b_N^+)^{n_N}}{\sqrt{n_N!}} \right) |0_0 0_1 0_2 \ldots 0_N\rangle = \prod_{i=0}^{N} \otimes \frac{(b_i^+)^{n_i}}{\sqrt{n_i!}} |0_i\rangle \quad (16)$$

The bosonic network particle position is represented in sequence by the lower index *i* on the $n_i$ $\{i \mid 0, 1, 2, \ldots, N\}$. The particle number at a given position *i* is represented by the integer value assumed by the $n_i, \{i \mid 0, 1, 2, \cdots, \infty\}$. The most general state for the bosonic network will include arbitrary superpositions of energy eigenstates for each position. Assuming a maximum finite energy eigenstate M, this most general state may therefore be represented as:

$$|\Psi_j\rangle = \sum_{j=1}^{N} \alpha_j |n_{0j} n_{1j} n_{2j} \ldots n_{Nj}\rangle = \sum_{j=1}^{M^{(N+1)}} \alpha_j \prod_{i=0}^{N} \otimes \frac{(b_i^+)^{n_{ij}}}{\sqrt{n_{ij}!}} |0_{ij}\rangle \quad (17)$$

Having used a unitary change-of-basis matrix $\mathbf{U}^+$ to transform $\mathbf{b}$, the vector of mutually interacting creation operators, into $\mathbf{b}'$, a vector of non-interacting operators. Eq. (17) may therefore be rewritten in the transformed basis as well via $\mathbf{U}$. The result is a new set of multi-boson state vectors each ket of which is a different mixture of the original kets. By then explicitly operating upon the vacuum state—which is the same in all bases—we obtain the transformed kets expressed as $|n'_0 n'_1 n'_2 \ldots n'_N\rangle$. Note that *M* will become *M'*: While the number of quasi-boson positions equals *N*, the same as the number of boson positions, the mixture of operators may lead to a different (overall) maximum eigenvalue.

The Hamiltonian in the new basis is simply Eq. (17), which may be expanded in more transparent form as

$$H_D = \sum_{i=0}^{N} \omega_{ii} b'_i b'^+_i \quad (18)$$

As there is no interaction among the quasi-bosons, all terms contain only the product of the creation and annihilation operators for that position (meaning no terms in which only a single creation or annihilation operator appears for any one position). This product is the number operator, and any time-developed ket may be expressed by replacing the number operator by the appropriate eigenvalue:

$$|n'_0 n'_1 n'_2 \ldots n'_N\rangle \rightarrow e^{ib'^+_0 b'_0 \omega t} |n'_0\rangle e^{ib'^+_1 b'_1 \omega t} |n'_1\rangle e^{ib'^+_2 b'_2 \omega t} |n'_2\rangle \ldots e^{ib'^+_N b'_N \omega t} |n'_N\rangle = \prod_{i=0}^{N} e^{in_i \omega t} |n'_0 n'_1 n'_2 \ldots n'_N\rangle \quad (19)$$

The full time-developed state vector for network in the transformed basis thus becomes:

$$|\Psi'\rangle = \sum_{j=1}^{M'^{(N+1)}} \prod_{i=0}^{N} \alpha'_j e^{in'_i \omega t} |n'_0 n'_1 n'_2 \ldots n'_N\rangle \qquad (20)$$

Each term in the sum has its own propagator factor. When the $|n'_i\rangle$ are transformed back to the original basis, these factors get mixed, but the solution, however complex, is exact.

Now that we have reviewed the method (useful in its own right) for obtaining a basis in which a network of (both bosonic and fermionic) oscillators are decoupled, we are ready to use it toward the end of devising a supersymmetric network.

The actual bosonic oscillator Hamiltonian, with ground state energy $+\frac{1}{2}$ (in suitable units) is

$$H_B = \frac{\hbar \omega_B}{2}(b^+ b + \mathbf{1}) \qquad (21)$$

The spin-1/2 fermionic Hamiltonian, with ground state energy $-\frac{1}{2}$ (in suitable units) is:

$$H_F = \frac{\hbar \omega_F}{2}(f^+ f - \mathbf{1}) \qquad (22)$$

The defining characteristic of a supersymmetric Hamiltonian is that it may be expressed as the square of some "supercharge" $Q = Q(b^+,b,f^+,f)$ such that the ground-state energy is exactly zero.. This means that if we presume a multiparticle system-plus-environment wherein each $\omega_{B_i} = \omega_{F_i}$ and $N_F = N_B = N$ (so that the total number of particle positions = 2$N$), the simple sum of their individual Hamiltonians forms a supersymmetric oscillator network Hamiltonian, $H_{SUSY}$ *that happens to be the natural square of a Q*:

$$H_B + H_F = H_{SUSY} = \frac{\hbar}{2}(b_i^+ b_i + f_i^+ f_i) = \frac{\hbar}{2}(b_i^+ f_i + b_i f_i^+)^2 = \frac{\hbar}{2} Q_i^2 \qquad (23)$$

From the mathematical point of view, we may look at this $H_{SUSY}$ as though it describes the energy structure of 2$N$ non-interacting particles, half labeled by $b_i$ and half by $f_i$ without further regard to what species of particle they happen to be. Supersymmetry allows bosons to be treated on a equal footing as fermions. The $+\mathbf{1}$ identity operator which we previously ignored we may now reintroduce since it is canceled by the $-\mathbf{1}$ identity operator in the fermionic Hamiltonian. The Landau splitting of electron energies in a magnetic field is the prototypical physical exemplar, within specified limits of precision (e.g., degree of symmetry brokenness—how above or below exactly zero is the real ground state), and in the limit that the interaction between the external field and the magnetic field induced by electron spin is inconsequentially small.

This means that any physical system displaying a supersymmetrical structure may be diagonalized such that there are no interactions among the quasi-bosons and the quasi-fermions. The advantage for quantum information processing follows from the distinctive action of the operators $Q$ and $Q^+$. Since $Q = Q^+ = (b^+ f + bf^+)$; and given the relations:

| | | | |
|---|---|---|---|
| $b\|0_B\rangle = \emptyset$ | $b\|1_B\rangle = \|0_B\rangle$ | $b^+\|0_B\rangle = \|1_B\rangle$ | $b^+\|1_B\rangle = \sqrt{2}\|2_B\rangle$ |
| $f\|0_F\rangle = \emptyset$ | $f\|1_F\rangle = \|0_F\rangle$ | $f^+\|0_F\rangle = \|1_F\rangle$ | $f^+\|1_F\rangle = \emptyset$ |

it follows that:

$$Q(|0_B\rangle \otimes |1_F\rangle) = Q^+(|0_B\rangle \otimes |1_F\rangle) = (b^+f + bf^+)|01\rangle = (b^+f)|01\rangle + bf^+|01\rangle = |10\rangle \quad (24)$$

$$\therefore Q|01\rangle = Q^+|01\rangle = |10\rangle;$$

and that:

$$Q|10\rangle = Q^+|10\rangle = |01\rangle \quad (25)$$

Furthermore:

$$Q^+Q|01\rangle = QQ^+|01\rangle = Q^2|01\rangle = H_{SUSY}|01\rangle = E_{SUSY}|01\rangle = 1|01\rangle \quad (26)$$

and

$$Q^+Q|10\rangle = QQ^+|10\rangle = Q^2|10\rangle = H_{SUSY}|10\rangle = E_{SUSY}|10\rangle = 1|10\rangle \quad (27)$$

In other words, the state $|01\rangle$ = 0 bosons and 1 spin $+\frac{1}{2}$ fermion, and the state $|10\rangle$ = 1 boson and 1 spin $-\frac{1}{2}$ fermion are the two degenerate eigenstates of $H_{SUSY}$. It therefore follows that:

$$Q\left(\frac{|01\rangle \pm |10\rangle}{\sqrt{2}}\right) = Q^+\left(\frac{|01\rangle \pm |10\rangle}{\sqrt{2}}\right) = Q^+Q\left(\pm\left(\frac{|01\rangle \pm |10\rangle}{\sqrt{2}}\right)\right) = Q^2\left(\pm\left(\frac{|01\rangle \pm |10\rangle}{\sqrt{2}}\right)\right) \quad (28)$$

where the states $|01\rangle \pm |10\rangle$ are two-particle entangled states between non-identical particles.

As the plus/minus signs outside the inner brackets Eq. (28) indicate only a common phase, they may be ignored. This does not imply, however, that the symmetrical and the anti-symmetrical states combining $|01\rangle$ and $|10\rangle$ are stationary under the influence of a general 50:50::boson:fermion environment. As noted above, the spin (fermionic) state $\frac{1}{\sqrt{2}}(|01\rangle - |10\rangle)$ is "effectively a two-particle eigenstate of the interaction Hamiltonian for each of two [spin-½] particles and an ensemble of many other [such] particles." Such a state is 100% resistant to decoherence by an external environment, regardless of size or (spin) basis of either action or representation. By contrast, the spin (fermionic) state $\frac{1}{\sqrt{2}}(|01\rangle + |10\rangle)$ is 100% resistant to decoherence only vis-à-vis an environment acting in its own basis, but subject to spin-flipping by an environment acting in an orthogonal basis, and in which orthogonal basis its representation would be completely decoherent.

The two-particle boson-fermion states $|10\rangle$ and $|01\rangle$ are eigenstates of the supersymmetric Hamiltonian, $Q^+Q = H_{SUSY}$. But (in contrast to fermionic states) the symmetrical *and* the anti-symmetrical superpositions of the two-particle boson-fermion states are eigenstates of $H_{SUSY}$—and not only of $H_{SUSY}$ but of the supersymmetric creation and annihilation operators $Q$ and $Q+$ as well. These unique relations, definitional of supersymmetry, suggest that both the symmetric and the anti-symmetric boson-fermion superpositions of Eq. (28) should have be fully resistant to decoherence. As basic a phenomenon as Landau splitting may therefore provide a useful substrate for relatively decoherence-free quantum information processing.

However, the "balance of forces" introduced by antisymmetry in the case of spins, and by supersymmetry in the case of boson-fermion pairs, requires that the phase angle between components of the two-particle superposition remains stationary. Intuitively, therefore we expect that for a composite boson-fermion system to display the necessary supersym-metry, the fermionic coupling strength to the environment and the bosonic coupling strength to the environment must be the same within any given pair (as we indeed just posited). In fact, supersymmetry allows us to relax this constraint considerably. Let us examine first the case when the paired constants are identical.

Suppose we have only four particles—two boson-fermion pairs, one pair being the "system," the other the "environment." To conveniently obtain the time developed state, the Bogoliubov transformation can be carried out in block-diagonal form (separating the boson-boson and fermion-fermion interactions before diagonalizing further). The target system supersymmetric "qubit" in the entangled states of Eq. (28), may be expressed in the original basis as:

$$\frac{\left(b_0^+ \otimes \mathbf{1}_1\right)}{\sqrt{2}}|0_B 0_F\rangle + \frac{\left(\mathbf{1}_0 \otimes f_1^+\right)}{\sqrt{2}}|0_B 0_F\rangle = \frac{1}{\sqrt{2}}\left(b_0^+ + f_1^+\right)|00\rangle \tag{29}$$

As before, we transform the bosonic and fermionic creation operators, but separately:

$$\mathbf{b}' = \mathbf{U}_B^+ \mathbf{b}; \mathbf{b}'^+ = \mathbf{b}^+ \mathbf{U}_B; \mathbf{f}' = \mathbf{U}_F^+ \mathbf{f}; \mathbf{f}'^+ = \mathbf{f}^+ \mathbf{U}_F \tag{30}$$

The propagator is obtained directly in the diagonalized basis and the inverse transformation is applied to obtain the time-developed state of the supersymmetric qubit:

$$\frac{\left(e^{\frac{i}{2}\Omega_{B_0}t} + e^{\frac{i}{2}\Omega_{B_1}t}\right)\left(b_0^+ \otimes \mathbf{1}_1\right)}{\sqrt{2}}|0_B 0_F\rangle \pm \frac{\left(e^{\frac{i}{2}\Omega_{B_0}t} + e^{\frac{i}{2}\Omega_{B_1}t}\right)\left(\mathbf{1}_0 \otimes f_1^+\right)}{\sqrt{2}}|0_B 0_F\rangle = \tag{31}$$

$$\frac{1}{\sqrt{2}}\left(e^{\frac{i}{2}\Omega_{B_0}t} + e^{\frac{i}{2}\Omega_{B_1}t}\right)|10\rangle \pm \frac{1}{\sqrt{2}}\left(e^{\frac{i}{2}\Omega_{F_0}t} + e^{\frac{i}{2}\Omega_{F_1}t}\right)|01\rangle$$

If the $\Omega_{B_i} = \Omega_{F_i}$, then the exponential factors introduced by the time-development will be common to both terms in the superposition and there will be no decoherence.

What if there are interactions between the bosons and fermions? This would mean that the Hamiltonian would contain terms such as:

$$\omega_{ij} b_i^+ f_j + \omega_{ij}^* b_i f_j^+ \tag{32}$$

Thirty years ago, Nicolai [4] described a one-dimensional supersymmetric oscillator network with composite SUSY supercharge operators $Q$ and $Q^+$ defined by:

$$Q_n \equiv \sqrt{\frac{\hbar}{2}} \sum_{i=1}^{N} \left(b_i^+ f_{i+n} + b_i f_{i+n}^+\right) \tag{33}$$

Note that every possible combination of one boson and one fermion is represented in any $Q_n$. Furthermore, for all $n$,

$$H_{SUSY} = \frac{Q_n^2}{2} = \frac{\hbar}{2} \sum_{i=1}^{N} |\omega_i|^2 \left(b_i^+ b_i + f_i^+ f_i\right) \tag{34}$$

which is simply the Hamiltonian for N non-interacting bosons and N non-interacting fermions—an expression which we may obtain by transforming the basis of representation for arbitrary number of mutually interacting bosons plus an equal number of arbitrarily interacting fermions.

But now suppose that the argument of Eq. (33) expressed not a term in any of the supercharges $Q_n$, but a term in its own square, the supersymmetric Hamiltonian, $H_{SUSY}$, to which it is not in general equal, of course. If this supposition were so, then we could re-express the fermion-boson interactions as a mere duplication of the independent boson-boson plus fermion-fermion interactions. In that case, any state that was an eigenstate (coherent state) of the supercharge would automatically be an eigenstate (coherent state) of the Hamiltonian.

As just remarked, this is not generally so: The eigenvalue of an operator is not automatically the eigenvalue of the square ($\mathbf{O}^+\mathbf{O} = \mathbf{O}^2 = \mathbf{O}$) of the operator—unless the operator in question is, for example (*inter alia*), the density matrix for a pure state; and if a superposed pure state, then a decoherence-free state.

While this is not generally so, it is specifically so when the state in question is precisely *either* $\frac{1}{\sqrt{2}}(|0_B 1_F\rangle + |1_B 0_F\rangle)$ *or* $\frac{1}{\sqrt{2}}(|0_B 1_F\rangle - |1_B 0_F\rangle)$. In other words, if we have a quite general Hamiltonian for a matched number of fermions and bosons, with arbitrary mutual interactions among all particles; and if we are willing to transform our basis appropriately, then as a qubit, *both* such states will automatically form a DFS.

One approach to evade decoherence in quantum information processing is to seek quantum states that are from the start entangled in such a way as preserve coherence, and to use these entangled states as a computational basis. Anti-symmetric singlet states are one example; a perhaps surprising second instance are analogous subspaces within a super-symmetric system, where there exist two such states. The ubiquity of boson-fermion multi-particles may suggest both new ways of understanding decoherence and resistance to it, as well as new physical substrates for qubits.

The author gratefully acknowledges Erich Poppitz's many contributions to this paper and the support of Steven Girvin.